# A High-efficient Battery Charging System for Electric Vehicle

Zexuan Li [1,2] *

[1] Department of Mathematics, Physics and Electrical Engineering, Northumbria University, Newcastle, United Kingdom
[2] Nanyang Technological University, Singapore
*lize0010@e.ntu.edu.sg

**Abstract:** Nowadays, automobile is facing the trend of electrification. Lithium-ion batteries is widely used as their power supplies. Lithium-ion battery has complex characteristics, as a result, Lithium-ion battery needs optimal charging strategies to make sure it is charged safely and efficiently. This paper focuses on development of a high-efficient charging method for lithium-ion battery. To test different charging strategies, the electric vehicle charging system consisting of a dual active bridge DC-DC converter and a Thevenin battery model is implemented. Multistage constant current charging (MSCC) and multistage constant current reflex charging (MSCC with reflex charging) were proposed. Compared with the traditional constant voltage constant current (CC-CV) charging method, MSCC can reduce 12% of the charging time and 1.1% of the battery loss; MSCC with reflex charging has a 10.45% and a 1.54% reduction of charging time and battery loss separately.

## 1. Introduction

In recent years, due to environmental protection and climate changes, the automobile is facing the trend of electrification. The UK government decided to put a ban on new petrol and diesel cars in the UK from 2030. Many car manufacturers have announced their new cars powered by electricity. Compared with conventional vehicles, EVs are mainly powered by batteries which are charged by electricity generated from solar, wind, and other renewable sources, so that they do not produce greenhouse gases and other toxic gases while running.

Nowadays, EVs are mainly using lithium-ion batteries as their power supplies because Li-ion battery has the advantages like long life, high power density, and high working voltage [1, 2]. One of the biggest disadvantages of EVs is that it takes a much longer time to fully charge the batteries compared with vehicles consuming fossil fuels. Moreover, the large amount of heat is produced during the charging period which leads to a giant loss of electricity.

As there won't have a significant improvement in the features of batteries. Different advanced battery charging strategies have come out to shorten the charging time. However, Lithium-ion batteries may become explosive and dangerous when it is charged with high power. As a result, Lithium-ion battery needs optimal charging strategies to make sure it works safely and efficiently.

### 1.1. Literature Review

After the lithium-ion battery is invented, consumers thought it is inconvenient to recharge these high-density batteries, because it takes a long time. The reason why the charging time is too long is that a dedicated charging strategy must make sure it meets strict safety requirements and life cycle requirements. To reduce the charging time of the lithium-ion battery, an ultra-fast charging method has been developed for over 20 years.

In 2006, P.H.L. Notten, J.H.G. Op het Veld and J.R.G. van Beek from Philips Research Laboratories and Eindhoven University of Technology proposed a "boostcharging(BC)" recharging algorithm. Based on the characteristic that close-to-empty batteries can be recharged with high voltage for a short time without introducing negative effects to the battery, they carried out experiments to prove the feasibility of this algorithm [3].

Experiments based on two types of battery, cylindrical (US18500, Sony) and prismatic LP423048(Philips) Li-ion batteries were used. Compared to their boostcharging method with the conventional CC-CV charging, a short period of constant voltage charging is added before the ordinary CC-CV charge stage. The voltage of the BC period $V_B^{max}$ is a little bit higher than that of the CV period $V_{CV}^{max}$ ((e.g., $V_B^{max} = 4.3V$ and $V_{CV} = 4.2V$). The experiment revealed that a large amount of charge can be stored during the BC period. More than 50% storage can be accomplished within 10 minutes, whist 30% storage can be charged in 5 min under boost charging conditions [3].

Jun Li, Edward Murphy, Jack Winnick, Paul A. Kohl from Georgia Institute of Technology investigated pulse charging, another method to achieve fast charging [3]. In their paper, impedance spectroscopy and cyclic voltammogram were used to compare the cycling characteristics and electrochemical behavior of lithium-ion batteries charged by the pulse charging method with that of the constant current constant voltage (CC-CV) charging method.

They found that it takes approximately 1 h to fully charge the battery by the pulse charging method at the 1 C pulse charging rate, while dc charging process required around 3.5 h [4]. Compared with the conventional constant dc charging method, the batteries charged by pulse have a larger discharge capacity at the same charge-discharge rate.

However, the experiment shows that the interfacial resistances of batteries cycled by pulse charging are slightly larger than that of the CC-CV charging method which can lead to an increase in interfacial charge-transfer resistances and surface film resistances. But this increase in the interfacial resistance did not directly relate to the fade of discharge capacity [4].

In conclusion, pulse charging that has short discharge pulses and short relaxation periods can reduce concentration and polarization which increases the power transfer rate so that less charging time is needed to fully recharge. Moreover, discharge capacity becomes larger and cycle life becomes longer due to the active material utilization [4].

In [5], a multistage CC-CV charging strategy was proposed. A larger capacity can be obtained by reducing the charging current when the terminal voltage reaches the cut-off voltage. Experiments showed that compared with CC-CV charging and two-stage CC-CV charging, multistage CC-CV charging has the shortest charging time and largest charging capacities at -10℃, 0℃, and 25℃.

Two issues must be settled when designing the multistage current changing strategy. First is determining the timing when current stages switch, another issue is setting the appropriate value of the current at each stage.

In [6], fuzzy control was used to design a five-step charging algorithm. according to the battery temperature, the appropriate charging current is set by the controller. Y. Liu and Y. Luo implemented the Taguchi method to obtain the optimal number of the current of each stage.

## 2. Methods
### 2.1. Dual Active Bridge DC-DC Converter (DAB)

To change the value of the voltage which the AC-DC converter transfers from the power grids to the primary side of the charging system, a DC-DC converter is needed before the battery package. In this project dual active bridge DC-DC converter is used. DAB is composed of two full-bridge DC/AC converters with 4 transistors and an isolation high-frequency transformer (See Fig.1).

The leakage inductor of the transformer is important to this system. By controlling the phase shift between the primary side and secondary side to change the direction and magnitude of the current of the inductor, the power flow direction can be shifted [7].

The DAB has several advantages, making it suitable to be implemented as an electric vehicle charging circuit. Firstly, DAB is an isolated DC-DC converter. Compared to the non-isolated DC-DC converter, DAB has the advantage of having isolation between the primary side and the secondary side. This makes the charging circuit safer and improves its reliability. Secondly, by changing the phase shift angle, the converter is able to transmit power forward and backward, which is essential because, in the reflex charging process, the battery needs to discharge. Moreover, compared with Forward and Flyback DC-DC converters, DAB is capable to transmit high power to achieve rapid charging [8].

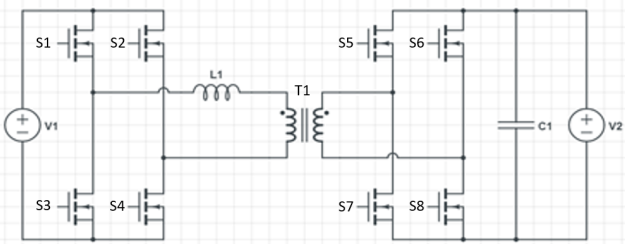

**Fig. 1.** *Circuit of DAB Converter*

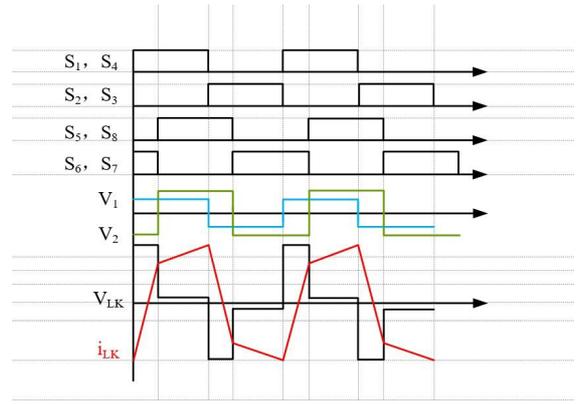

**Fig. 2.** *Operation waveform of DAB*

DAB has several function modes. The most common one is the single-phase shift mode (See Fig.2). $S_n$ is the control signal to the transistors. The primary side (input side) and secondary side (output side) operate in a 50% duty cycle, So the output voltage of each bridge $V_{AB}$ is a square wave. There is a phase shift between the primary side and the secondary side. $S_5 S_8$ start later than $S_1 S_4$. Moreover, to prevent shoot through, there is a dead time after the conducting transistors stop working and before the other two transistors turn on. $V_{LK}$ is the voltage difference between the primary side output voltage and the reflected voltage. $I_{LK}$ is the current goes through the inductor.

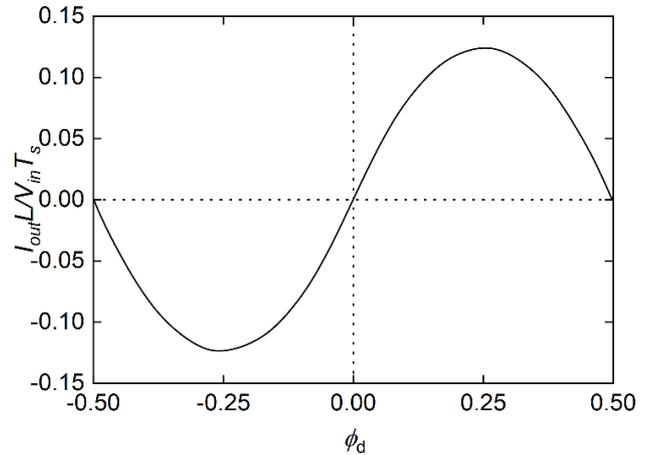

**Fig. 3.** *Phase Shift Control*

The relationship between the input voltage and output current can be described as:

$$I_{out} = \frac{Q_2 + Q_A - Q_B}{0.5T_s} = \frac{V_{in}T_s\varphi_d|1-2\varphi_d|}{nL} \quad (1)$$

$$P_{in} = \frac{nv_1v_2}{2fL}d(1-d) \quad (2)$$

$$i_2 = \frac{nv_1}{2fL}d(1-d) \quad (3)$$

As shown in Fig.3, by changing the phase shift angle of DAB, power transmit direction and value can be controlled. The parameter of phase shift angle $\varphi_d$ needs to be controlled in the range from -0.25 to 0.25.

The loss of the converter is mainly caused by the MOSFETs, the high-frequency transformer, the resonant inductor, and the wire resistance [9]. In this paper, only MOSFETs' loss is considered.

The loss of MOSFETs includes the switching loss and conduction loss, which are shown in (4)(5)(6).

$$P_{on} = 2 \times (R_{DS\,on\,H1} + K^2 R_{DS\,on\,H2}) \times I_{rms}^2 \quad (4)$$
$$P_{off-on} = \frac{1}{6} V_{DS}(t_1) \times I_D(t_2) \times t_r \times f_s \quad (5)$$
$$P_{on-off} = \frac{1}{6} V_{DS}(t_2) \times I_D(t_1) \times t_r \times f_s \quad (6)$$

### 2.2. Thevenin Battery Model

There are several types of Lithium-ion battery models. To simulate the transient state, the polarization effect and internal resistance need to be considered. Thevenin model is one of the most widely used models which can simulate the transient state.

The simplest Thevenin model is the first-order Thevenin model(See Fig.3). This model is composed of an open circuit voltage source($U_{oc}$), an ohmic resistance($R_o$), and an RC pair($C_{Th}$, $R_{Th}$) which can simulate the capacitance effect between two battery plates and contact resistance. By adding an RC pair to the model, the transient phenomena can be simulated [10, 11].

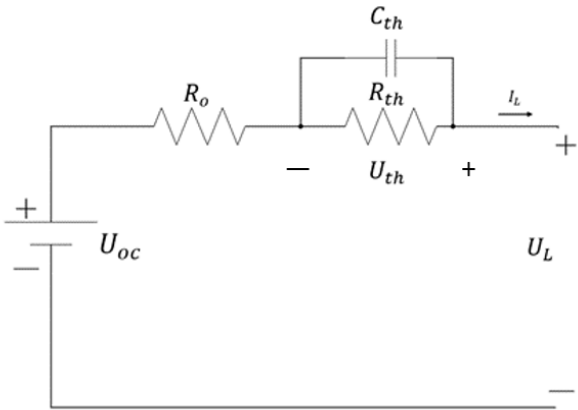

*Fig. 4. First-order Thevenin model*

The accuracy of the model can be improved by considering that the state of charge(SoC) of the battery can influence other parameters of the battery.

If the simulation needs higher precision, a second RC pair can be added. The second-order Thevenin model has a larger time constant so that it can describe the long-term transient effect better. The higher-order Thevenin model can have better accuracy, but more computation cost is also needed[9].

The mathematic model of the first-order Thevenin model and the loss of battery can be described as:

$$C_{Th}\frac{dU_{Th}}{dt} + \frac{U_{Th}}{R_{Th}} = I_L \quad (7)$$
$$U_{oc} + I_L R_o + U_{Th} = U_L \quad (8)$$
$$SoC = \frac{1}{Ah}\int_0^t I_L \, dt + SoC(0) \quad (9)$$

$$P_{loss} = I_L^2 R_o + \frac{U_{Th}^2}{R_{Th}} \quad (10)$$

### 2.3. Battery charging strategy

As for the existing charging method, traditional constant current-constant voltage(CC-CV) charging is the most widely used one because of its simplicity and easy implementation. In the recent 10 years, reflex charging, also, and multistage constant current charging were two of the most popular strategies because they can reduce the concentration and polarization of the battery to largen the capacity and prolong the life cycle and charge the battery faster[12].

Different charging strategies have different advantages and drawbacks. For example, constant voltage charging can protect the battery from the high current at the high-SoC stage but because of the small current, the charging speed is quite slow. At the same time, constant current charging can charge the battery at a high speed, but a large current can make the battery overheat and shorten the lifecycle [12].

To settle this problem, different charging strategies can be combined. The whole charging period is divided into several stages. A controller is needed to decide which charging method should be implemented, and the specific figures of each charging stage.

In this project, three charging strategy is going to be tested. The first one is multistage constant current charging, another one is multistage constant current reflex charging which combines MSCC with reflex charging strategy, these two advanced charging methods will be compared with traditional CC-CV charging.

*2.3.1 Constant Current Constant voltage Charging:* The constant current-constant voltage(CC-CV) charging strategy is one of the most developed and widely used lithium-ion battery charging methods, because of the advantages of simplicity and easy implementation[12]. In the progress of CC-CV charging, at first, a constant current is used to charge the battery, the constant current is usually the maximum current that battery can withstand. After battery voltage reaches a pre-set maximum charging voltage($V_{pre-set}$), the charging voltage level off at this number, while the charging current is correspondingly reduced exponentially. The charging progress stops when the charging current reduces to a predetermined small current.

The flowchart of CC-CV charging is shown in Fig.5. The charging current and charging voltage need to be carefully decided to prevent the battery from overcharge. The main disadvantage of CC-CV charging is that it takes a long time to fully charge the battery because of the small charging current of the constant voltage charging stage [12]. The newly developed charging methods mostly aim to reduce the charging time by replacing the constant voltage charging stage with advanced control methods.

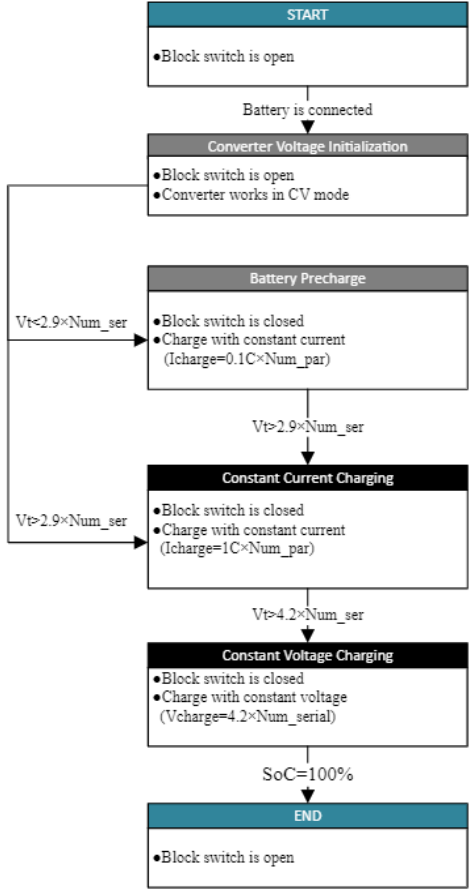

*Fig. 5. Flowchart of CC-CV*

*2.3.2 Multistage Constant Current Charging (MSCC):* Multistage constant current charging is an algorithm that divides constant current charging into several stages. The charging currents of each stage decrease while the SoC number increases because of the characteristic of lithium-ion batteries [13].

Two important features, the number of charging stages and the current value of each stage need to be decided when designing the algorithm [14]. The charging stages' number decides the control accuracy of the charging process.

With more charging stages, at each stage battery can be charged with the most suitable current. However, more stages need a more complex control system. Both computation cost and precise control need to be considered when deciding the charging stages' number. Among different multistage charging system designs, the five-stage structure is most widely used because of its simplicity and accuracy [14].

The flowchart of a five-stage MSCC is shown in Fig.6. The main multistage charging stage consists of 5 small stages which have different charging current values. When the terminal voltage of the battery reaches the pre-set value (in this case 4.2V), it turns to the next charging stage.

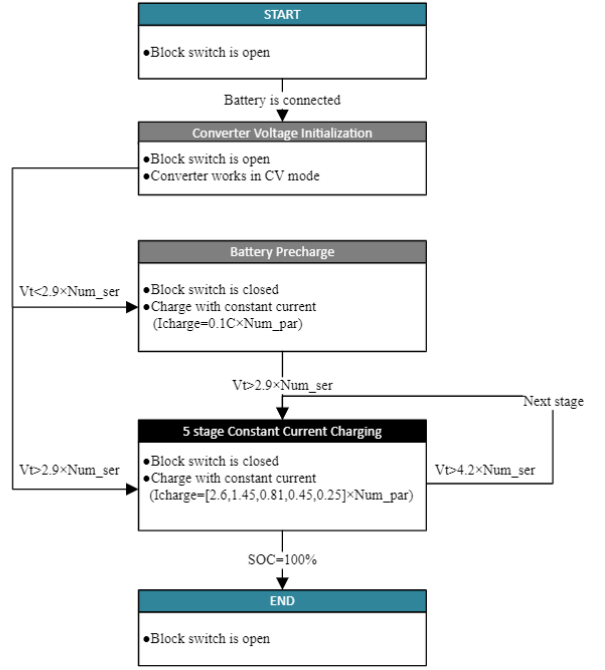

*Fig. 6. Flowchart of MSCC*

As for the charging current value and length of each stage, they are usually determined by experiments [15, 16] or mathematic modelling [14]. In this paper, a method is introduced to determine the current value.

The first-order Thevenin model can be equivalent to a terminal voltage ($V_t$) and a capacitor ($C_1$) and a resistor $R_1$ in serial connection.

$V_o$ is the initial terminal voltage, so that the value of terminal voltage ($V_t$) and voltage across capacitor ($V_{C_1}$) and voltage across resistor ($V_{R_1}$) is shown below:

$$V_t = V_{C_1} + V_{R_1} \quad (11)$$

$$V_{C_1} = \frac{1}{C_b} \int_{t_0}^{t} I(\tau) d\tau + V_o \quad (12)$$

$$V_{R_1} = IR_1 \quad (13)$$

$$V_t = \frac{1}{C_1} \int_{t_0}^{t} I(\tau) d\tau + V_o + V_{R_1} \quad (14)$$

Because $I(\tau)$ is constant at each stage, the time of each stage ($\Delta t_x$) and total charging time can be calculated in (19).

$$\Delta t = \frac{1}{I} (V_t - V_o - V_{R_1}) \times C_1 \quad (15)$$

$$T = \Delta t_1 + \Delta t_2 + \Delta t_3 + \Delta t_4 + \Delta t_5 \quad (16)$$

$$V_o^x = V_t - I_{x-1} \times R_1 \quad x = 2,3,4,5 \quad (17)$$

$$V_{R_1}^x = I_x \times R_1 \quad x = 1,2,3,4,5 \quad (18)$$

$$\Delta t_x = \frac{1}{I_x} (V_t - V_o^x - V_{R_1}^x) \times C_1 \quad (19)$$

Assuming there are three charging stages, the charging time can be written and by calculating the derivative of time and letting it equal to zero, the regulation determining the best middle-stage charging current is shown in (22).

$$T = \Delta t_1 + \Delta t_2 + \Delta t_3 \quad (20)$$

$$\frac{dT}{dI_2} = \frac{-I_1 R_1 C_1}{I_2^2} + \frac{R_1 C_1}{I_3} = 0 \quad (21)$$

$$I_2 = \sqrt{I_1 I_3} \quad (22)$$

As for five-stage charging the middle stage charging currents also obey the same regulation. As shown in (23)(24)(25).

$$I_3 = \sqrt{I_1 I_5} \quad (23)$$

$$I_4 = \sqrt{I_3 I_5} \quad (24)$$

$$I_2 = \sqrt{I_1 I_3} \quad (25)$$

*2.3.3 MSCC with Reflex charging:* reflex charging also known as bipolar pulse charging is an advanced charging method that can charge the battery fast and efficiently. Reflex charging consists of positive and negative pulses. Pulsed current can help to reduce the concentration polarization and interface resistance [17, 18]. So that larger power can be transmitted, the charging time can be reduced. Moreover, the battery can have a long lifetime. Negative pulsed current is also important because it has the function of depolarization, reducing the lithium dendrites inside the battery [19].

Due to the huge advantages, reflex charging has, a new charging method is proposed by combining reflex charging with MSCC. The flowchart of the proposed MSCC with reflex charging is shown in Fig. 7. Reflex charging is added to the last period of every charging stage to reduce the polarization effect. Compared with MSCC, when the terminal voltage reaches the pre-set value, the charging pattern of MSCC with reflex charging turns to reflex charging. The absolute value of the charging current of reflex charging is the same as the current of stages of MSCC.

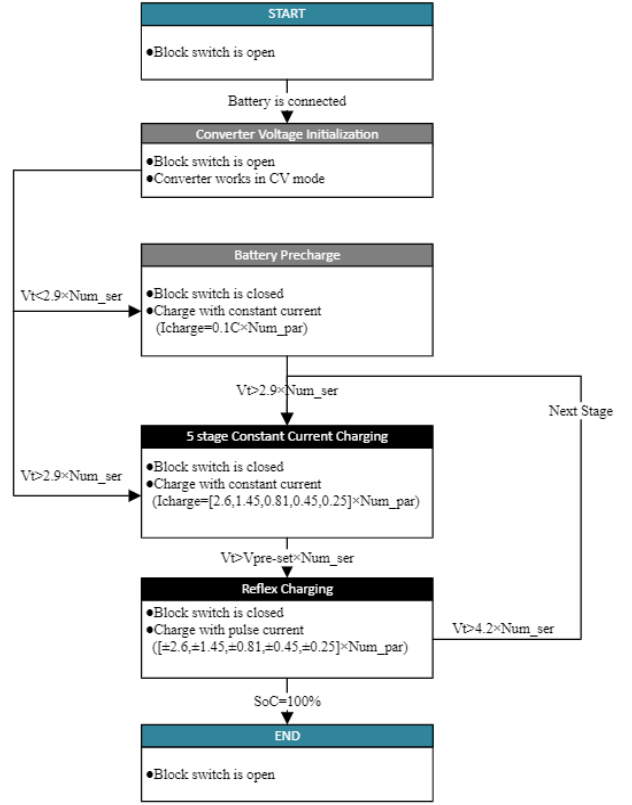

**Fig. 7.** *Flowchart of MSCC with Reflex Charging*

## 3. Simulation

The charging system is built in the MATLAB Simulink. The diagram of the charging system is shown in Fig.8. The system includes a dual active bridge DC-DC converter, a first-order Thevenin battery model, and a charging control system.

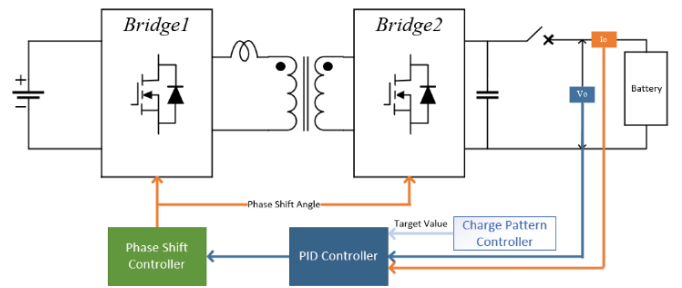

**Fig. 8.** *Charging System Diagram*

The dual active bridge converter is responsible to transmit power to the lithium-ion battery and control the charging current and voltage. The parameters of DAB are shown in Table 1.

| Table 1 Parameters of DAB | |
|---|---|
| Switch Frequency | 20kHz |
| Leakage Inductance | 15μH |
| Supply Side Voltage | 200V |
| Second Side Voltage | 150V |
| Second Side Capacitor | 10mF |

As for the battery model, Panasonic 18650B is used to build a 35*35 battery pack. The first-order Thevenin model can simulate the transient state precisely. The parameters of the battery model are obtained by conducting a hybrid pulse power characterization (HPPC) test. The parameter of a single battery cell model is shown in Fig. 9 and Table.2. In this paper, the temperature of the battery model is not considered.

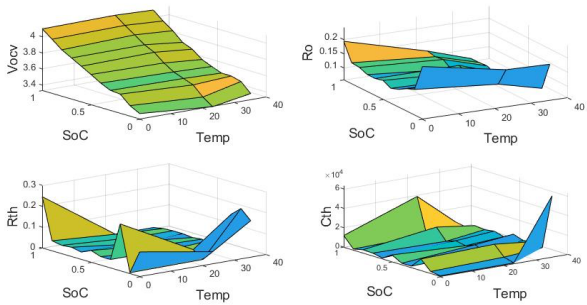

**Fig. 9.** *Parameters of Panasonic 18650B Battery* [20]

| Table 2 Parameters of Panasonic 18650B Battery Pack [20] | |
|---|---|
| Nominal Voltage | 3.6V |
| Maximum Charging voltage | 4.2V |
| Battery Capacity | 2600mAh |
| Standard Charging Current | 1C |
| Batteries in parallel | 35 |
| Batteries in serial | 35 |
| Initial SoC | 20% |

The whole system is controlled by closed-loop control. To make the output current and voltage of the DAB converter maintain a constant value, closed-loop discrete PID controllers are implemented. The input of the controller is the value of output voltage and current. After processing by the PID algorithm, the controller output the phase shift angle to the DAB phase shift controller, so that DAB can produce constant voltage constant current. By combining constant voltage control and constant current control together, different proposed charging strategies can be achieved. The built charging system in MATLAB Simulink is shown in Fig.10.

## 4. Results and Analysis

The DAB under single-phase control is tested in MATLAB Simulink. The waveforms of input output voltage ($V_1, V_2$) of DAB converter and the current and voltage of the leakage inductor ($I_L, V_L$) is shown in Fig.11. From the figure, it can be observed that there is a time delay between $V_1$ and $V_2$ causing the voltage across the leakage inductor. The DAB works properly under single-phase control.

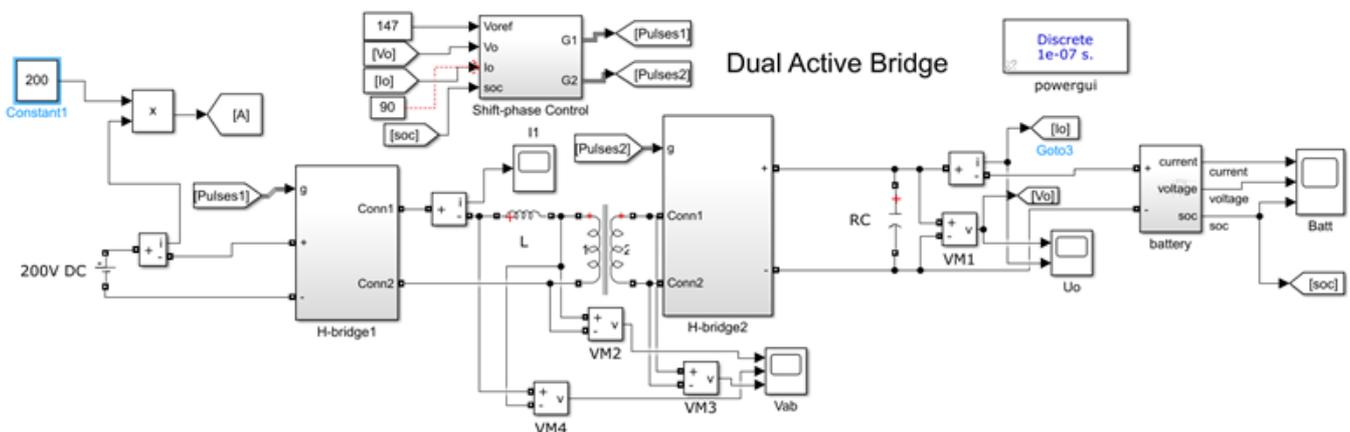

**Fig. 10.** *The Circuit in MATLAB Simulink*

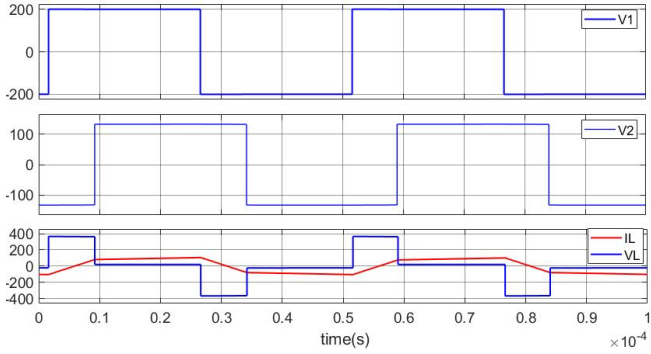
*Fig. 11.  Waveforms of DAB*

In Fig.12, the charging system charges the battery with constant current first, and the terminal voltage of the battery goes up. After it reaches the pre-set value, the charging system turns to the constant voltage charging.

Fig.13 and Fig.14 show that the charging system charges the battery according to the proposed charging strategy separately. The charging process consists of 5 constant current charging stages. Compared with MSCC, MSCC with reflex charging has the reflex charging stage

**Table 3** Charging Time and Loss

|  | CC-CV | MSCC | MSCC with Reflex Charging |
|---|---|---|---|
| Charging time | 1.244h | 1.093h | 1.114h |
| Battery Loss | 0.454kWh | 0.449kWh | 0.447kWh |
| Converter Loss | 0.139kWh | 0.159kWh | 0.1708kWh |
| Total Loss | 0.593kWh | 0.608kWh | 0.6178kWh |

The constant current constant voltage charging algorithm, multistage constant current charging, and multistage constant current reflex charging are implemented in the Simulink charging system model. The waveforms of each charging method produced by MATLAB Simulink are shown in Fig.12, Fig.13, and Fig.14, each figure contains three important parameters, charging current, the terminal voltage of the battery model, and SoC value.

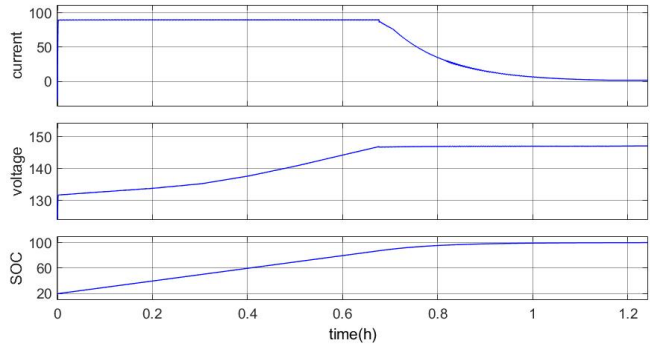
*Fig. 12.  Waveforms of CC-CV*

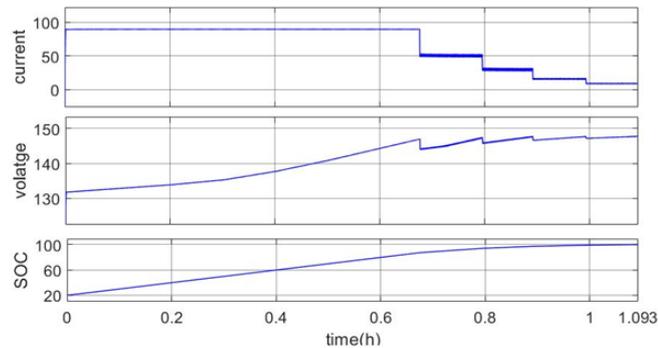
*Fig. 13.  Waveforms of MSCC*

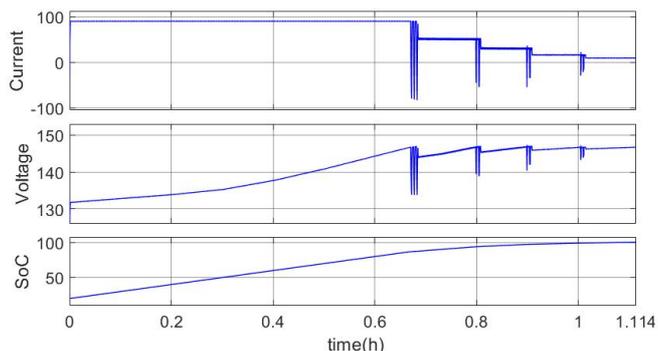
*Fig. 14.  Waveforms of MSCC with Reflex Charging*

Based on the simulation, the charging time and loss of the model are calculated and compared with each other. As shown in Table.3.

It can be explicitly seen from the result that the charging time of MSCC is 12% shorter than conventional CC-CV charging, which is shortest among these three tested charging strategies, and it can also reduce the loss of battery by 1.1%. However, the loss on the converter of MSCC and total loss is larger than CC-CV charging.

MSCC with reflex charging also can reduce the charging time by 10.45%, but still needs more charging time than MSCC. Moreover, MSCC with reflex charging method produces the smallest battery loss among all three charging strategies (1.54% less than CC-CV). But MSCC with reflex charging also produces the largest converter loss and largest energy loss in total.

Despite more loss produced by MSCC with reflex charging, the bipolar pulse of MSCC with reflex charging can prevent the precipitation of lithium crystals and prolong the lifetime of the battery. These advantages can compensate for the drawback of more loss.

In a nutshell, the simulation results demonstrate that the control system works effectively, and both two proposed charging methods have a better performance compared with conventional CC-CV charging, MSCC can charge the battery fastest and MSCC with reflex charging produces the smallest battery loss.

## 5.  Conclusion

This paper proposed a 5-stage multistage constant current charging method and a 5-stage multistage constant current with reflex charging strategy. A dual active bridge DC-DC converter and its control system were developed to achieve dynamic control of different charging methods. The first-order Thevenin battery model was built to simulate the characteristic of lithium-ion batteries. Finally, the MSCC charging method and MSCC with reflex charging method were simulated in the MATLAB Simulink model and simulation results were obtained. It was found that MSCC and MSCC with reflex charging can shorten the charging time by 12% and 10.45% separately compared with conventional CC-CV charging. MSCC and MSCC with reflex charging can also reduce 1.1% and 1.54% of the charging loss on the battery than the CC-CV method. Both two advanced charging methods show better performance than CC-CV charging method.

In future research work, to make the simulation closer to reality, the element of temperature needs to take into consideration. Apart from the simulation work, tests need to conduct on real 18650 lithium-ion battery packages, using testing equipment in the lab to get the real data on charging time and loss. Moreover, discharge tests and measurements of the polarization effect under the microscope are also needed to test proposed charging strategies.